\documentclass[]{spie}  %>>> use for US letter paper
\usepackage{amsmath,amsfonts,amssymb}
\usepackage{graphicx}
\usepackage[colorlinks=true, allcolors=blue]{hyperref}
\title{A 10.24-GHz-wide digital spectrometer array system for LMT-FINER: system design and laboratory performance verification}

\author[a]{Masato Hagimoto}
\author[a]{Akio Taniguchi}
\author[a]{Yoichi Tamura}
\author[a]{Norika Okauchi}
\author[b]{Hiroaki Kawamoto}
\author[c]{Taku Nakajima}
\author[c]{Takumi Hikosaka}
\author[b]{Kenichi Harada}
\author[b]{Toru Taniguchi}
\author[d]{Takeshi Kamazaki}
\author[e]{Takeshi Sakai}
\author[f]{Kunihiko Tanaka}
\author[d,a]{Ryohei Kawabe}
\affil[a]{Department of Physics, Graduate School of Science, Nagoya University, Furocho, Chikusa-ku, Nagoya, Aichi 464-8602, Japan}
\affil[b]{Engineering Department, Elecs Industry Co., Ltd., 1‑22‑23 Takatsu, Kawasaki, Kanagawa 213‑0014, Japan}
\affil[c]{Institute for Space-Earth Environmental Research, Nagoya University, Furocho, Chikusa-ku, Nagoya, Aichi 464-8601, Japan}
\affil[d]{National Astronomical Observatory of Japan, 2-21-1 Osawa, Mitaka, Tokyo 181-8588, Japan}
\affil[e]{Graduate School of Informatics and Engineering, The University of Electro-Communications, Chofu, Tokyo 182-8585, Japan}
\affil[f]{Department of Physics, Faculty of Science and Technology, Keio University, 3-14-1 Hiyoshi, Yokohama, Kanagawa 223-8522, Japan}

\authorinfo{Further author information: (Send correspondence to Masato Hagimoto): \\E-mail: hagimoto@a.phys.nagoya-u.ac.jp, Telephone: +81-(52)-789-2840}

\begin{document} 
\maketitle

\begin{abstract}
For efficient spectroscopic redshift identification of early galaxies in the northern hemisphere, we aim to combine the Large Millimeter Telescope (LMT) with a wide-band heterodyne receiver, FINER, which will cover radio frequencies of $120$--$360~\mathrm{GHz}$ and offer a $3$--$21~\mathrm{GHz}$ intermediate frequency (IF) per sideband and polarization. 
To take full advantage of such wide IFs, we present a novel 10.24-GHz-wide digital spectrometer, DRS4 (Elecs Industry Co., Ltd.). 
It incorporates $20.48~\mathrm{Gsps}$ samplers with an FPGA-based digital signal processing module. 
To mitigate the noise contamination from the image sideband, it is equipped with a digital sideband separation function to improve the sideband rejection up to $25~\mathrm{dB}$. 
Laboratory performance evaluations show that it exhibits an Allan time of at least $\sim100~\mathrm{s}$ and a total power dynamic range of at least $7~\mathrm{dB}$. 
These results demonstrate its capability of instantaneously wide-band spectroscopy toward high-redshift galaxies with position-switching observations.  
\end{abstract}

\keywords{Spectrometer, wideband, (sub)millimeter wavelength, digital signal processing}

\section{Introduction}
\label{sec:introduction}

When massive galaxies emerge and how common they are in the epoch of reionization and beyond (i.e., the age of the universe less than 600 million years, or the pre-reionization era) is one of the most fundamental questions in astronomy.
Recent discoveries of the excessive massive galaxy candidates at $z > 8$ made by \textit{Hubble Space Telescope} and \textit{James Webb Space Telescope} suggest an extremely high efficiency of galaxy assembly in the earliest Universe, while it is still unclear what drives the rapid growth of the massive population.
This arises from the difficulty in confirming galaxy ``candidates'' in such era, i.e., identifying an object as a galaxy and determining its physical properties with spectroscopy of atomic or molecular emission lines.
For $z > 8$, millimeter and submillimeter-wave spectroscopic observations of redshifted far-infrared emission lines such as  \textsc{[C\,ii]}$~158~\mu\mathrm{m}$ and \textsc{[O\,iii]}$~88~\mu\mathrm{m}$ should be a key to address the question.
They, however, are very faint (expected flux density to be $\sim$5~mJy; \textsc{[O\,iii]}$~88~\mu\mathrm{m}$ of a Lyman break galaxy with $H_{160} = 25.0$~AB at $z = 10$) and require wide frequency coverage ($\sim 30$~GHz) to confirm the spectroscopic redshift due to a large uncertainty of the photometric redshift ($\Delta z_{\mathrm{phot}} \sim 1$).
Sensitive and efficient spectroscopic instruments with sufficient atmospheric conditions to reach the furthest universe, even including ALMA, have been very limited yet.

For this sake, we have been developing Far-Infrared Nebular Emission Receiver, FINER (Tamura et al. 2024, submitted to Proc. SPIE), for the Large Millimeter Telescope (LMT\cite{Hughes2010, Hughes2020}) in M\'{e}xico.
FINER consists of Band 4+5 and 6+7 dual-polarization sideband-separating superconductor-insulator-superconductor (SIS) mixer receivers to continuously cover a wide range in radio frequencies from 120--360~GHz.
Utilizing high critical current density ($J_{c}$) SIS mixers\cite{Kojima2017, Kojima2020} developed for the ALMA Wideband Sensitivity Upgrade (WSU\cite{Carpenter2022}), FINER offers an instantaneous intermediate frequency (IF) of 3--21 GHz per sideband per polarization, which is $\sim$5 times wider than ALMA.
Combining the 50-m dish of LMT (40\% of the light-collecting area of ALMA) with similar atmospheric transmission to the ALMA site, LMT-FINER will achieve the most sensitive spectral-scanning capability among (sub)millimeter single-dish telescopes in the northern hemisphere.

To take full advantage of such wideband IF signals in the backend, development of a wideband ($\gtrsim 10$~GHz) digital spectrometer is essential to efficiently achieve continuous and homogeneous spectral scans.
With the recent progress in direct signal sampling by high-speed analog-to-digital converters (ADCs) and digital signal processing (DSP) by field-programmable gate arrays (FPGAs) and graphics processing units (GPUs), fast Fourier transform (FFT)-type (or FX-type) digital spectrometers have achieved several GHz of instantaneous bandwidths (e.g., XFFTS\cite{Klein2012};  OCTAD-S\cite{Iwai2017}).
For LMT-FINER, 10~GHz or more instantaneous bandwidth is preferred to fully cover the four IFs of a receiver (product of two polarization modes and lower/upper sidebands, or LSB/USB) by eight boards.
For spectral scans with ground-based single-dish telescopes, atmospheric noises from an image sideband with lower atmospheric transmission should also be mitigated.
In addition to analog sideband separation by a sideband-separating receiver (10--15~dB), DSP may offer a capability of digital sideband separation (DSBS\cite{Finger2015, Rodriguez2018}) to compensate for the imbalance of complex gain (i.e., amplitude and phase) between the sidebands and to remove the image sideband signal digitally ($\gtrsim 20$~dB).
Spectrometers with such capabilities, however, have not yet been available for actual observations.

\begin{figure}
    \centering
    \includegraphics[width=\textwidth]{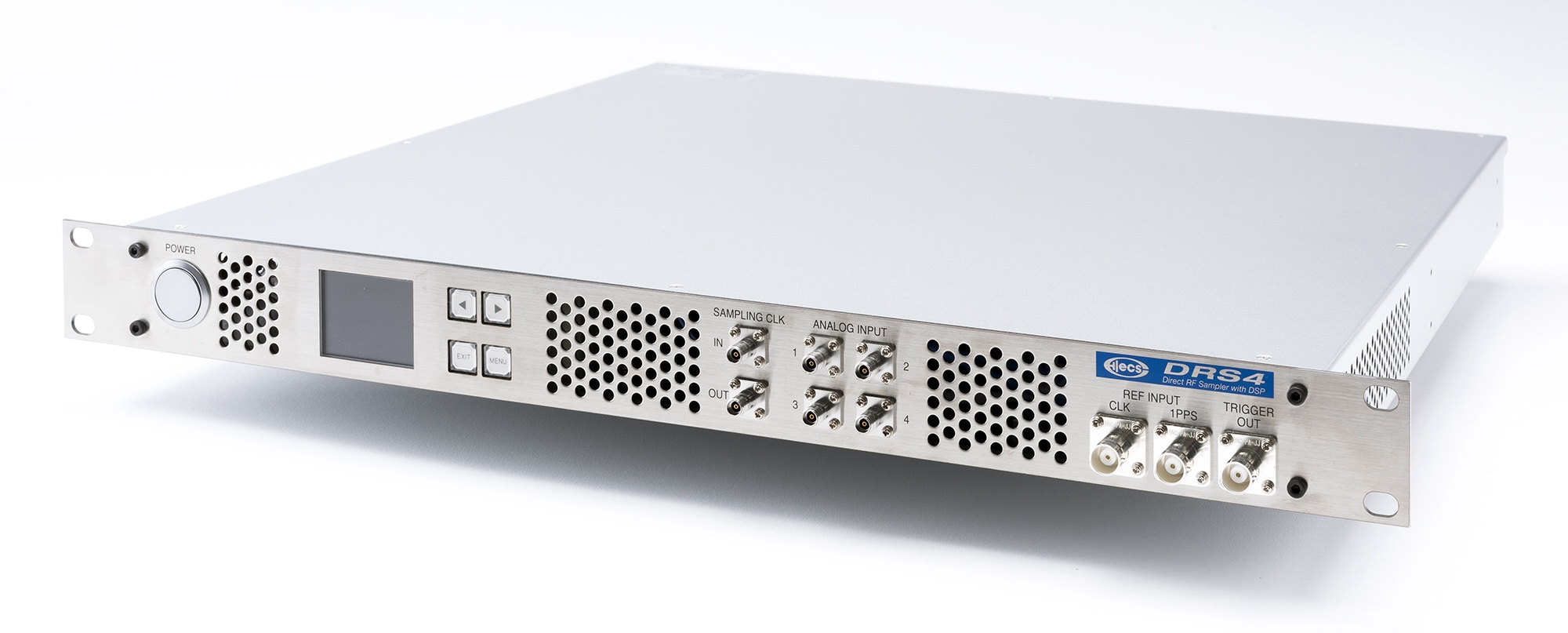}
    \caption{
        Front view of DRS4 customized for LMT-FINER.
        The LCD displays real-time power spectra and is used for device configuration together with four buttons on the side.
        The \texttt{ANALOG INPUT} port pairs 1/2 and 3/4 correspond to LSB/USB IF inputs for DC--10.24~GHz and 10.24--20.48~GHz, respectively.
        The \texttt{SAMPLING CLK IN/OUT} port receives/sends an external 20.48~GHz reference signal from the DRS4 Sampling Clock Distributor (Elecs Industry Co., Ltd.) for operating the four 20.48-Gsps ADCs.
        The \texttt{REF INPUT 1PPS} port receives an external 1~PPS signal for setting timestamps of the power spectra.
        Note that absolute date and time information should be given by an NTP server through the Ethernet port on the back panel.
        The Ethernet port is also used to output (multicast) the power spectrum data as VDIF.
        Other ports are not used in this model.
    }
    \label{fig:drs4-front}
\end{figure}

Here we present the design and laboratory performance verification of the digital spectrometer array system for LMT-FINER.
It is an array of new 10.24-GHz-wide digital spectrometers with DSBS functions, Direct RF Sampler with DSP 4$^{\mathrm{th}}$ Generation (DRS4; Elecs Industry Co., Ltd.; Figure~\ref{fig:drs4-front}), each of which has four analog inputs to directly samples the signals at 20.48~Gsps with a good analog transmission up to 20.48 GHz\footnote{We hereafter refer to the DRS4 model customized for LMT-FINER simply as DRS4.}.
This offers the capability of single-polarization, two-sideband spectroscopy by a single DRS4, i.e., the baseband (DC--10.24~GHz) and the 10.24--20.48~GHz band with no secondary downconverters.
We have manufactured two DRS4s so far to meet the baseline specification of LMT-FINER.
In this manuscript, we describe the basic specifications of DRS4 and the DSBS implementation in Section~\ref{sec:drs4-a-10.24-ghz-wide-digital-spectrometer}, the results of the laboratory performance verification in Section~\ref{sec:laboratory-experiments}, and the conclusions with future prospects before and after the installation of FINER on LMT in Section~\ref{sec:conclusions-and-future-prospects}.
\section{DRS4: a 10.24-GHz-wide digital spectrometer}
\label{sec:drs4-a-10.24-ghz-wide-digital-spectrometer}

\begin{table}[ht]
    \caption{Basic specifications of DRS4 for LMT-FINER} 
    \label{tab:basic-specifications-of-drs4}
    \begin{center}       
    \begin{tabular}{|l|l|}
    \hline
    \rule[-1ex]{0pt}{3.5ex} Item & Specification\\
    \hline\hline
    \rule[-1ex]{0pt}{3.5ex} Number of analog inputs & 4 (two pairs of LSB/USB)\\
    \hline
    \rule[-1ex]{0pt}{3.5ex} Optimal analog input power (DC--10.24~GHz) & -11 -- -9~dBm\\
    \hline
    \rule[-1ex]{0pt}{3.5ex} Optimal analog input power (10.24--20.48~GHz) & -8 -- -6~dBm\\
    \hline
    \rule[-1ex]{0pt}{3.5ex} Quantization levels of ADCs & 8 (3 bits)\\
    \hline
    \rule[-1ex]{0pt}{3.5ex} Sampling rate of ADCs & 20.48~GHz\\
    \hline
    \rule[-1ex]{0pt}{3.5ex} Instantaneous bandwidth per input & 10.24~GHz\\
    \hline
    \rule[-1ex]{0pt}{3.5ex} Number of FFT points & 1024\\
    \hline
    \rule[-1ex]{0pt}{3.5ex} Number of frequency channels & 512\\
    \hline
    \rule[-1ex]{0pt}{3.5ex} Frequency channel intervals & 20~MHz\\
    \hline
    \rule[-1ex]{0pt}{3.5ex} Window function & None (rectangle) / Hamming / Hanning\\
    \hline
    \rule[-1ex]{0pt}{3.5ex} Output data type & 512 32-bit floating points per input\\
    \hline
    \rule[-1ex]{0pt}{3.5ex} Output data dumping time & 100~ms / 200~ms / 500~ms / 1000~ms\\
    \hline
    \end{tabular}
    \end{center}
\end{table}

\subsection{Basic specification}
\label{sec:basec_spec}

DRS4 is a 10.24-GHz-wide FX-type digital spectrometer that has two pairs of LSB/USB analog inputs and digital outputs as auto-correlation with an optional output of cross-correlation between LSB and USB for the calibration of DSBS (Section~\ref{sec:calibration-of-the-digital-complex-gains}).
The basic specifications are summarized in Table~\ref{tab:basic-specifications-of-drs4}.

\begin{figure}
    \centering
    \includegraphics[width=\textwidth]{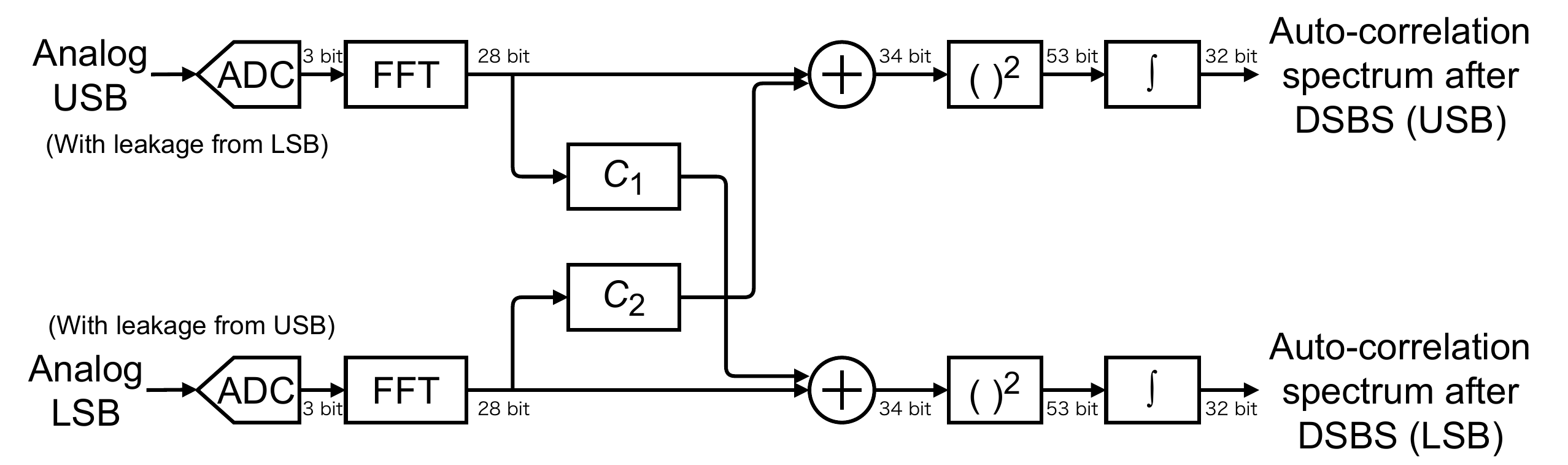}
    \caption{
        Block diagram of digital signal processing in an observation.
        Analog LSB/USB signals from a sideband-separating receiver (left side of the diagram) are analog-to-digital converted \fbox{ADC} and (fast) Fourier-transformed \fbox{FFT}to the frequency-domain complex data.
        They are then added to each other \fbox{+} after multiplying the digital complex gains \fbox{$C_{1}$} and \fbox{$C_{2}$} (i.e., DSBS).
        Finally auto-correlation \fbox{$()^{2}$} and time integration \fbox{$\int$} are performed to output LSB/USB power spectra after DSBS (right side of the diagram).
        Note that the numbers of bits indicate the bit-lengths of the data output from each block.
    }
    \label{fig:dsp-observation}
\end{figure}

Figure~\ref{fig:dsp-observation} shows the block diagram of DSP in each pair:
It consists of two 3-bit ADCs at a sampling rate of 20.48~Gsps and subsequent DSP modules made by FPGA boards to perform FFT and DSBS.
The 20.48~Gsps reference signal is externally given by the DRS4 Sampling Clock Distributor (Elecs Industry Co. Ltd.) for operating the ADCs.
The 1024 FFT points result in the number of frequency channels of 512 and frequency channel intervals of 20~MHz (20~km~s$^{-1}$ at 300~GHz)\footnote{For the observations with higher frequency resolutions, FINER also offers six XFFTS boards ($2^{15}$ frequency channels with frequency channel intervals of 88~kHz or 0.088~km~s$^{-1}$ at 300~GHz) and will switch to DRS4 as needed.}.
Two frequency-domain complex data are then added to each other after multiplying the complex digital gains $C_{1}$ and $C_{2}$ to compensate for the imbalance between two sidebands and to remove the possible image sideband signal (i.e., DSBS).
This similarly works to an analog 90-degree hybrid circuit but better separates two sidebands by optimizing $C_{1}$ and $C_{2}$ in calibration (Section~\ref{sec:calibration-of-the-digital-complex-gains} in details).

After DSBS, each complex data is auto-correlated, cast to 32-bit floating points, and integrated at a given data dumping time to obtain a single power spectrum as VDIF\cite{Whitney2009}.
The timestamp of each VDIF data is externally given by a network time protocol (NTP) server and a 1~pulse-per-second (PPS) signal.
The series of VDIF data are finally output (multicast) from the Ethernet port.
Data acquisition software (a binary and a Python script) is provided to allow users to get them in their local machine for further offline analysis.

\subsection{Calibration of the digital complex gains}
\label{sec:calibration-of-the-digital-complex-gains}

To enable DSBS in DRS4, the complex digital gains $C_{1}$ and $C_{2}$ (for each frequency channel) must be optimized by a calibration measurement of a reference signal such as continuous wave (CW) that only enters the signal sideband and possibly leaks into the image sideband.
According to previous work\cite{Rodriguez2018}, the optimal $C_{1}$ and $C_{2}$ are given by:
\begin{align}
    \hat{C}_{1}
        &= -\left.\frac{\mathrm{FT}[V_{\mathrm{LSB}}]}{\mathrm{FT}[V_{\mathrm{USB}}]}\right|_{\mathrm{Ref.\,signal\,in\,USB}}
        = -\left.\frac{\mathrm{FT}[V_{\mathrm{LSB}}] \cdot \mathrm{FT}[V_{\mathrm{USB}}]^{\ast}}{\mathrm{FT}[V_{\mathrm{USB}}] \cdot \mathrm{FT}[V_{\mathrm{USB}}]^{\ast}}\right|_{\mathrm{Ref.\,signal\,in\,USB}}\\
    \hat{C}_{2}
        &= -\left.\frac{\mathrm{FT}[V_{\mathrm{USB}}]}{\mathrm{FT}[V_{\mathrm{LSB}}]}\right|_{\mathrm{Ref.\,signal\,in\,LSB}}
        = -\left.\frac{\mathrm{FT}[V_{\mathrm{USB}}] \cdot \mathrm{FT}[V_{\mathrm{LSB}}]^{\ast}}{\mathrm{FT}[V_{\mathrm{LSB}}] \cdot \mathrm{FT}[V_{\mathrm{LSB}}]^{\ast}}\right|_{\mathrm{Ref.\,signal\,in\,LSB}}
\end{align}
where $V_{\mathrm{LSB/USB}}$ are LSB/USB outputs after analog-to-digital conversion, respectively, and $\mathrm{FT}[\cdot]$ performs (fast) Fourier transform on the signal.
The rightmost sides of the equations suggest that the measurements of both auto-correlation and cross-correlation of two sidebands enables to calculate $\hat{C}_{1}$ and $\hat{C}_{2}$.

For this purpose, DRS4 has another DSP mode for calibration to output auto/cross-correlation spectra at the same time.
Figure~\ref{fig:dsp-observation} shows the block diagram of it:
When calculating $\hat{C}_{1}$ and $\hat{C}_{2}$, $C_{1}$ and $C_{2}$ are initially set to be $1 + 0j$.
With a dedicated Python command, the auto/cross-correlation spectra are measured and written into separate text files.
The calculated $\hat{C}_{1}$ and $\hat{C}_{2}$ should be written in a text file and sent to DRS4 with another command to set as new values.
Note that in the measurements using CW, users should sweep its output frequency for all or several representative channel frequencies, i.e., need multiple calibration measurements.

\begin{figure}
    \centering
    \includegraphics[width=\textwidth]{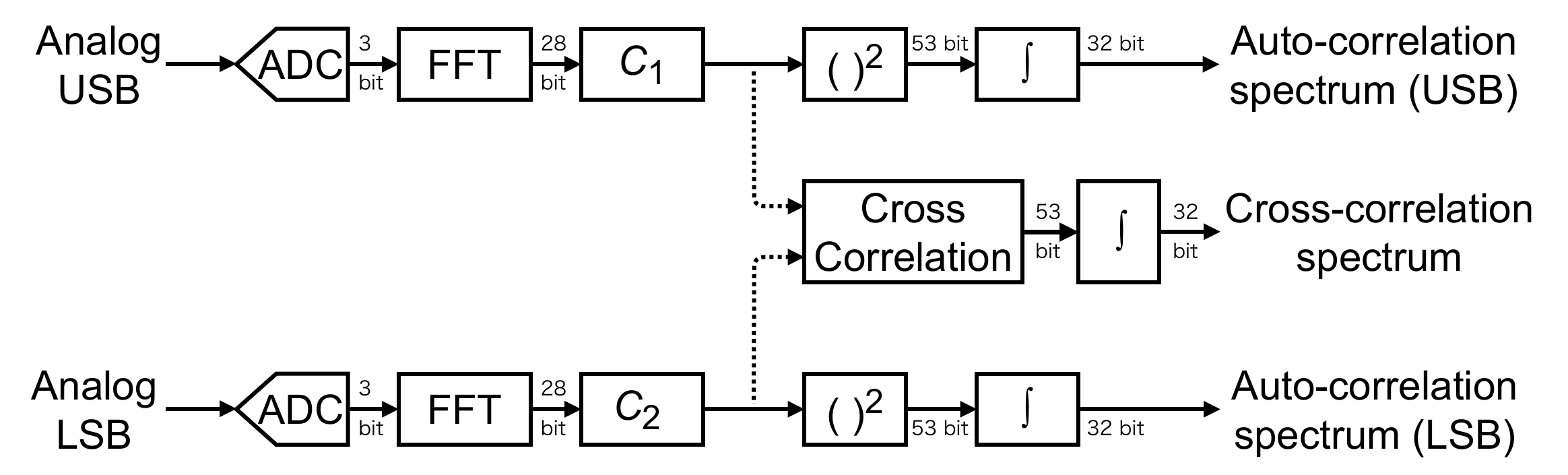}
    \caption{
        Block diagram of digital signal processing in a digital complex gain calibration.
        The components are almost same as Figure~\ref{fig:dsp-observation}, however, the major difference is that the two sideband signals are not added to each other before calculating auto/cross-correlation.
        The measured auto/cross-correlation spectrum are written as the comma-separated-value (CSV) format with channel frequency column.
        Note that the numbers of bits indicate the bit-lengths of the data output from each block.
    }
    \label{fig:dsp-calibration}
\end{figure}

\section{Laboratory experiments}
\label{sec:laboratory-experiments}
We evaluated the performance of DRS4 in laboratory experiments to confirm that we can use this to achieve our science goals.
Here we focused on three points; (i) frequency response function, (ii) total power and spectral line linearities, and (iii) time stability. 
This section summarizes the results of them.
\subsection{Response function}
\label{sec:lab_respncefunc}
The frequency response function is a response to the CW at a single frequency.
The spectrometer output is always obtained as the true emission line profile convolved with the frequency response function. 
Therefore, the actual frequency resolution is expressed as the full width at half maximum (FWHM) of it, not the frequency channel separation. 
DRS4 has three types of window functions (rectangle, Hann, Hamming), and users can select any of them. 
Here, we report the measurements for the rectangle window function of the finest resolution. 
In this case, the frequency response function is expected to be written in the form of $\mathrm{Sinc}^2(\tau \nu)$, where $\tau$ is the FFT segment length.
We expect the $\tau$ of $5.0\times10^{-8}~\mathrm{s}$ for our spectrometer, and thus the FWHM of it is calculated to be $17.72~\mathrm{MHz}$. 
\begin{figure}[ht]
    \centering
    \includegraphics[width=0.8\linewidth]{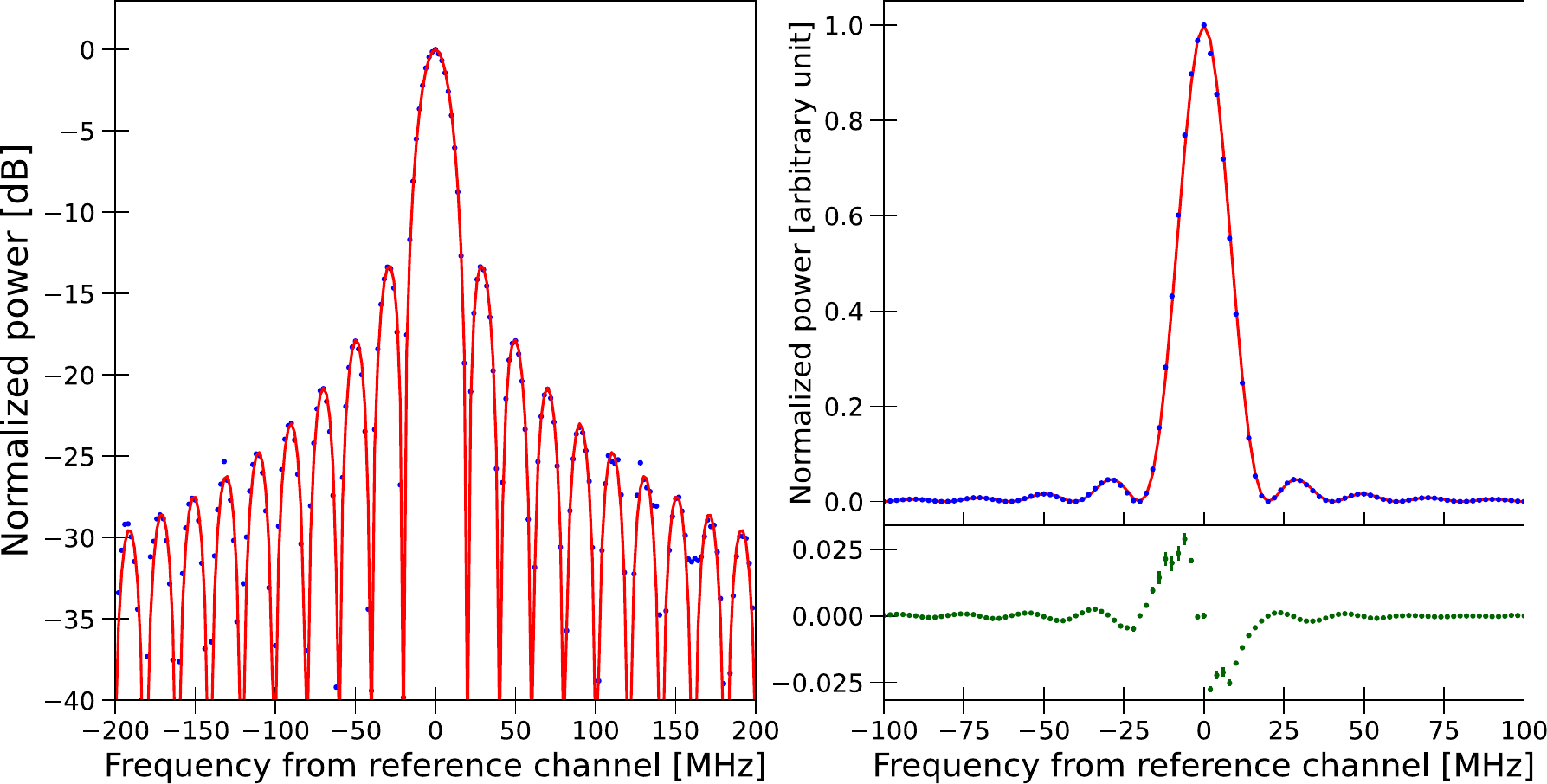}
    \caption{
    Frequency response function to the monochromatic wave input generated by a signal generator at $5.00~\mathrm{GHz}$ channel (blue points) with the best-fit curve of squared-Sinc function (red line) in log- (left panel) and linear-scale (right panel). Green points in the lower-right panel show residuals between measurements and best-fit curves (upper-right panel). We find that the FWHM of the best-fit curve is $0.7~\%$ larger than the theoretical one. 
    }
    \label{fig:responsefunc}
\end{figure}

Figure~\ref{fig:responsefunc} shows the frequency response curve at the $5.00~\mathrm{GHz}$ channel to the monochromatic wave input generated by a signal generator obtained by sweeping with $2~\mathrm{MHz}$ interval over a $400~\mathrm{MHz}$ range. 
We fit the measurements with $\mathrm{Sinc}^2(\tau\nu)$ to estimate the FWHM of the frequency response function.
The red line in Figure~\ref{fig:responsefunc} is the best-fit result and we confirm that the difference between the measurements and the best-fit function is small enough to reproduce the measurements by the theoretical predicted function.
% until at least the third sidelobe. 
We estimate the FWHM of $17.850\pm0.004~\mathrm{MHz}$, which is $0.7~\%$ larger than the theoretical value but is smaller than $20~\mathrm{MHz}$ of the channel separation.
In addition, the leakage from the neighboring channel is $\lesssim-10~\mathrm{dB}$. 
Therefore we consider that our spectrometer can resolve two neighboring channels sufficiently. 

\subsection{Linearity}
Ensuring linearity is important to convert the spectrometer output to the brightness temperature correctly because we intend to perform the intensity calibration using a room-temperature blackbody (or hot load) and sky temperature, i.e., R-sky calibration.
For this purpose, a wide dynamic range, the range where output power is proportional to input power, is required to correctly obtain both sky and room temperatures.
Here we verify the dynamic ranges of DRS4 in two aspects; total power and spectral line intensity.
\subsubsection{Total power linearity}
\label{sec:lab_total-power}
Total power measurement evaluates the system noise temperature and obtains the continuum flux densities. 
We used a noise signal generated with the noise source developed for the Millimetric Adaptive Optics (MAO\cite{Tamura2020}).
It generates the wide-band continuum signal in the frequency range of $17.4$--$23.6~\mathrm{GHz}$.
We made the signal narrower by combining a band-pass filter and used the folded component as $0.5$--$2.5~\mathrm{GHz}$.
We measured the input power of DRS4 with a power meter and the output power from DRS4 by integrating the frequency range above with changing the input power over a $12~\mathrm{dB}$ range by a variable attenuator with the step of $1~\mathrm{dB}$.
We did measurements with and without a chip attenuator of $6~\mathrm{dB}$ to investigate a wider input power range.
\begin{figure}[ht]
    \centering
    \includegraphics[width=0.8\linewidth]{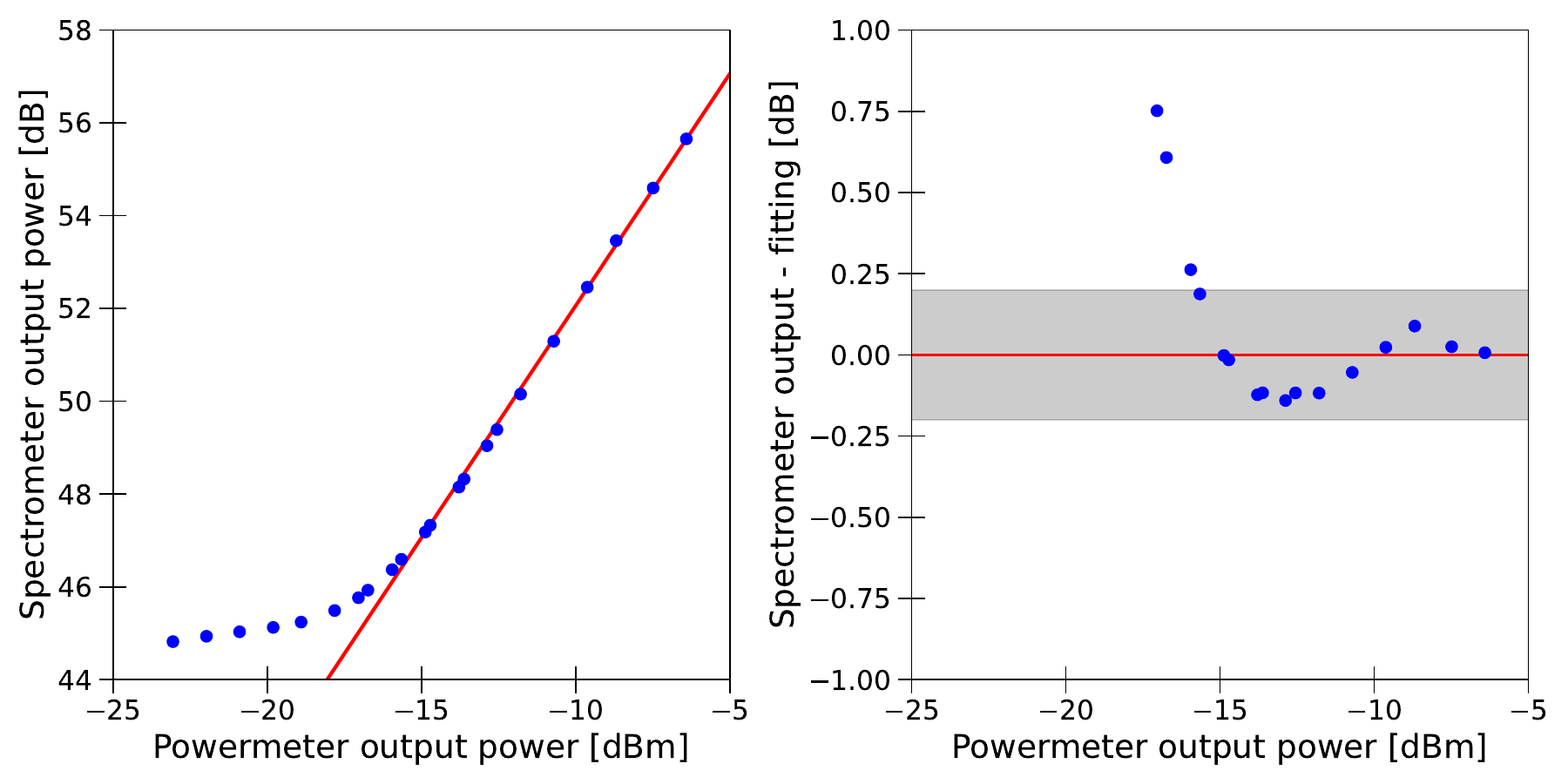}
    \caption{
    (Left) Spectrometer response for wide-band continuum input generated by noise source for MAO experiment (blue points) with the best-fit result of linear regression fitting with fixing a slope of unity (red line). (Right) Residual between measurement and the best-fit result (blue points). The gray-shaded region shows the range of $\pm0.20~\mathrm{dB}$, corresponding to our total power dynamic range definition. We find at least $\sim7~\mathrm{dB}$ of dynamic range for our spectrometer.
    }
    \label{fig:totalpower_linearity}
\end{figure}

The left panel of Figure~\ref{fig:totalpower_linearity} shows the result of our measurements with the best fit of the linear regression fitting of $\log P_\mathrm{out}=\log P_\mathrm{in}+b$.
We define the total power dynamic range as the range of input power where the difference between measurements and the best-fit line is within $\pm0.20~\mathrm{dB}$ (the gray-shaded region in the right panel of Figure~\ref{fig:totalpower_linearity}). 
Although we see non-linear responses in this range, we find at $\sim7~\mathrm{dB}$ of dynamic range for our spectrometer.
This satisfies the requirement from R-sky calibration expected for the LMT-FINER system.
In the future, we will make the calibration table for non-linearity of $\pm0.20~\mathrm{dB}$ to obtain the intensities more correctly.

\subsubsection{Spectral line linearity}
\label{sec:lab_on-off}
With DRS4, we aim to detect both the ``weak'' ($\sim 1~\mathrm{mK}$) emission line from galaxies at high redshift with LMT-FINER and ``strong'' ($\gtrsim10~\mathrm{K}$) emission line such as SiO maser for pointing calibration.
To evaluate performance in observations and obtain flux densities correctly, we need to have the dynamic range of spectral line linearity.
We made a mock spectral line signal embedded in a noise signal combining the noise source used in Section~\ref{sec:lab_total-power} with the monochromatic CW output generated by a signal generator at $18.48~\mathrm{GHz}$.
We observe this CW signal at $2.00~\mathrm{GHz}$ channel in DRS4 utilizing higher-order sampling mode.
To remove the time variance of the band-pass characteristics of our system, we switched the signal generator output on and off every three seconds.
We did measurements of spectrometer output power at $2.00~\mathrm{GHz}$ with and without a chip attenuator of $30~\mathrm{dB}$ to investigate a wider power range.
\begin{figure}[ht]
    \centering
    \includegraphics[width=0.8\linewidth]{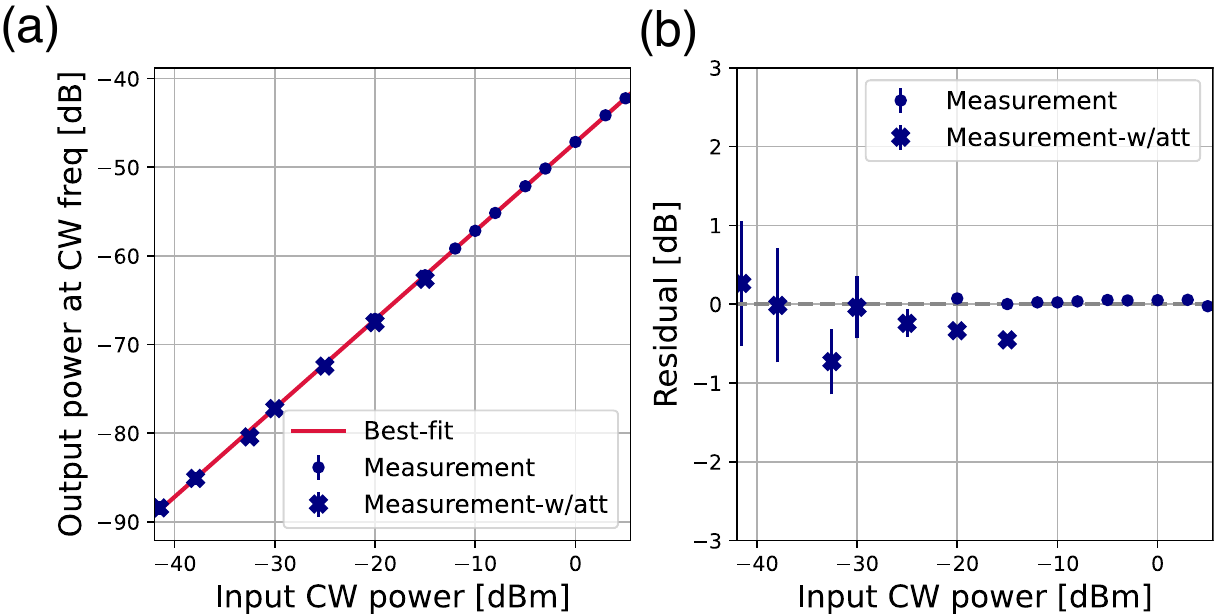}
    \caption{
    (Left) Spectrometer response of the single-frequency CW signal embedded in continuum at $2.00~\mathrm{GHz}$ (blue circles/crosses; without/with attenuators) with the best-fit result of liner regression fitting with fixing slope of unity (red line). (Right) Difference between the best-fit result and measurements in the left panel. We can find at least $\sim45~\mathrm{dB}$ of spectral line linearity though there are larger variations on the lower-power side than the higher-power side. This variation seems to be due to the insertion loss in attenuators. 
    }
    \label{fig:onoff_linearity}
\end{figure}

Figure~\ref{fig:onoff_linearity} shows the result of our measurements with the best fit of the linear regression fitting of $\log P_\mathrm{out}=\log P_\mathrm{in}+b$ and the differences between measurements and best fit.
We confirm DRS4 has at least $45~\mathrm{dB}$ of dynamic range for spectral line signal, and this satisfies our science requirement.
Note that we find residual offsets in the measured power with and without chip attenuators.
This is considered to be due to the frequency dependency of the insertion loss of the attenuators. 
Although it does not change our conclusions, in the future, we will evaluate and correct them to achieve more precise results.

\subsection{Time stability}
\label{sec:lab_Allanvariance}
\begin{figure}[ht]
    \centering
    \includegraphics[width=1.0\linewidth]{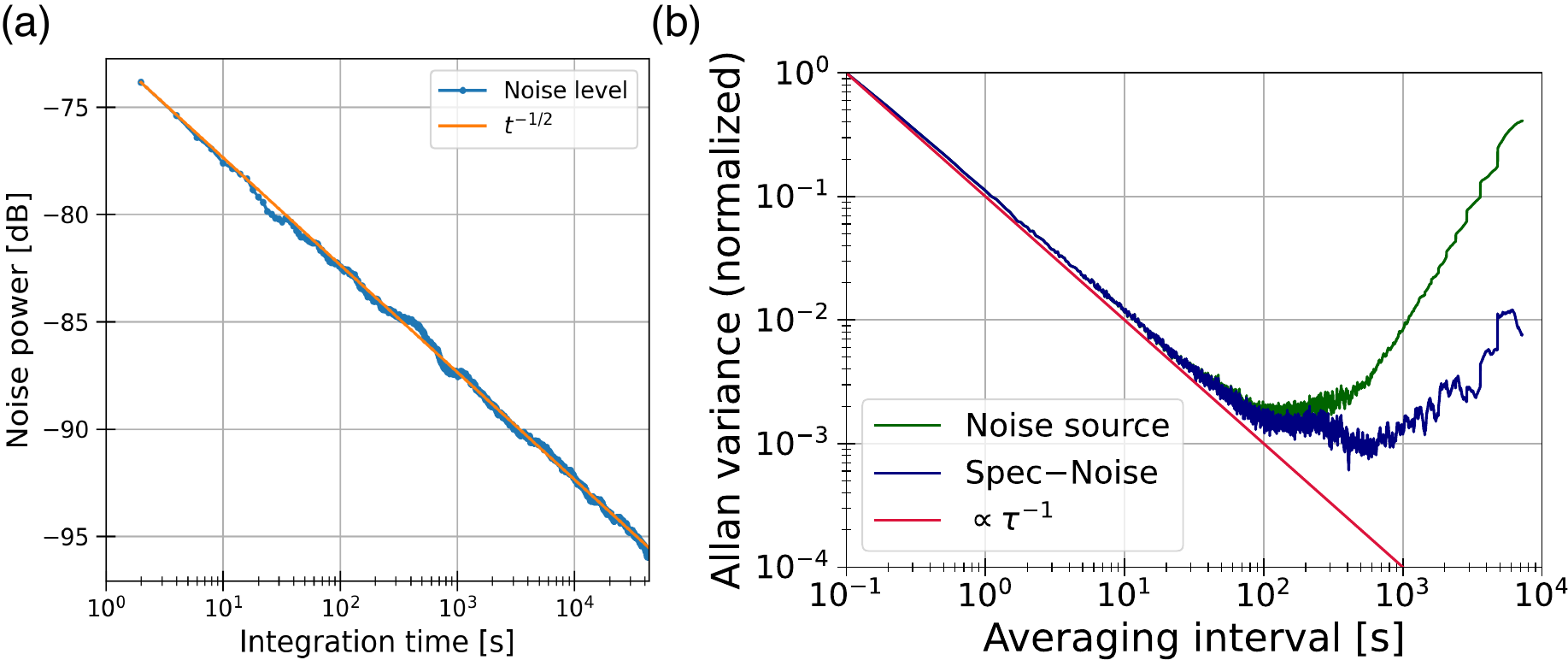}
    \caption{
    (Left) Noise level measurement for long integration (blue points) with theoretical prediction ($\propto t^{-1/2}$; orange line). Noise level decreases following $\propto t^{-1/2}$ until at least $43200~\mathrm{s}=12~\mathrm{hours}$, which meets our science requirement. (Right) Normalized time-based Allan variance of the noise source (green) and the spectrometer subtracted the contribution to the instability from the noise source (blue). The red line indicates the power law with the index of $-1$, which corresponds to the predicted relation toward white noise with frequency modulation. We can find the Allan time of $\sim100~\mathrm{s}$ for DRS4 but need to consider more carefully that there is still a contribution from the instability of the noise source.
    }
    \label{fig:Stability}
\end{figure}

% We aim to detect emission lines from bright galaxies at $z>8$ by integrating the spectrum for $\geq8~\mathrm{hours}$, and thus we need to demonstrate that our spectrometer can decrease the noise level for such a long time.
The root-mean-square noise level of the integrated spectra should be improved proportional to the inverse of the root of the observation time. 
Since we aim to integrate for a night to detect emission lines from bright galaxies at $z>8$, it should be improved at least for $\sim8~\mathrm{hours}$.
Here we measured the integration time dependence of noise level with on-off switching experiments.
We used the same signal of Section~\ref{sec:lab_on-off} and integrated for up to $43200~\mathrm{s}=12~\mathrm{hours}$. 
The left panel of Figure~\ref{fig:Stability} shows the result of noise level measurement (blue points) with the theoretical prediction (orange line). 
We find that the noise level decreases following $\propto t^{-1/2}$ for $12~\mathrm{hours}$.
This satisfies our science requirement.

% We need long-term integration to detect the astronomical signal from high-redshift galaxies.
We cannot obtain the spectra of on- and off-source positions simultaneously.
Therefore, the instability of a spectrometer can cause systematic errors, which results in the instability of spectral baselines.
Here we measured the time-based Allan variance with continuum signal.

To evaluate time stability, we calculated the time-based Allan variance $\sigma^2(\tau)$ for the total power of the continuum spectrum.
$\sigma^2(\tau)$ is defined as
\begin{equation}
    \sigma^2(\tau)=\frac{1}{2(N-1)}\sum^{N-1}_{i=1}\left(P_{i+1}-P_{i}\right)^2,
\end{equation}
where $P_i$ is the $i$th of $N$ fractional powers averaged over the sampling interval, $\tau$.
We used the same noise source in Section~\ref{sec:lab_total-power} and measured the time variation of total power with the power meter simultaneously as the reference. 
To remove the contribution to the instability of our spectrometer from the noise source, we simply subtract the power meter output from that of the spectrometer. 
The right panel of Figure~\ref{fig:Stability} shows the result of our measurements and the relation expected toward the white noise with frequency modulation ($\sigma^2(\tau)\propto t^{-1}$)\cite{Riley2008}.
We find that the $\sigma^2(\tau)$ of our spectrometer is proportional to $\tau^{-1}$ relation until $\sim100~\mathrm{s}$, which is shorter than the Allan time for other FX-type digital spectrometers (e.g., XFFTS\cite{Klein2012}, OCTAD-S\cite{Iwai2017}).
This shorter Allan time may be due to the incomplete subtraction of the noise source instability by our simple analysis above, and additional measurements with a stable noise source should be necessary for further study.
We should note that the current Allan time is longer than the typical time scale of the atmospheric variance ($\sim10~\mathrm{s}$) and thus still meets the required switching interval between on- and off-source positions.

\section{Conclusions and future prospects}
\label{sec:conclusions-and-future-prospects}
We have designed a spectrometer array system for the LMT-FINER project and evaluated a new 10.24-GHz-wide digital spectrometer developed as a component of our system. 
To identify the spectroscopic redshifts of high-redshift galaxies for as wide a frequency range as possible, we implement a function of the digital sideband separation in our spectrometer.
We evaluate the performance of DRS4 in laboratory experiments.
The results are summarized as follows;
\begin{enumerate}
    \item Frequency response function is well-fitted by $\mathrm{Sinc}^2(\tau x)$. The predicted FWHM is $0.7~\%$ higher than the expected one ($17.72~\mathrm{GHz}$), but this is enough small to separate between two adjacent frequency channels.
    \item We obtained a total power dynamic range of at least $7~\mathrm{dB}$, which is enough large to cover the R-sky calibration expected for science observations with LMT-FINER. In the future, we need to make the calibration table for non-linearity of $\pm0.20~\mathrm{dB}$ to obtain the intensities more correctly.
    \item We found the spectral line linearity over at least $45~\mathrm{dB}$ with on-off switching measurement of the embedded monochromatic CW signal into the noise signal. This enables us to detect both strong emissions for pointing observations and weak emissions from high-redshift galaxies with correct intensities.
    \item We confirm the noise level decreases following $\propto t^{-1/2}$ with integration time until $43200~\mathrm{s}$, which satisfies science requirement. The obtained time-based Allan time of $\sim 100~\mathrm{s}$ is shorter than other FX-type digital spectrometers, but this probably includes the instability of the noise source. We need to subtract their contribution to the instability of the spectrometer more correctly.
\end{enumerate}

We are now establishing the calibration method of the complex digital gains practically and demonstrating the improvement of the sideband rejection ratio (SRR) by DRS4.
As an initial result, we achieved $\sim7~\mathrm{dB}$ improvement compared to only analog sideband separation.
In the future, we aim to achieve $25~\mathrm{dB}$ of SRR as a whole system by combining DRS4 with the FINER receiver.

% DSBSについては、少なくとも7 dBの上昇を確認できた (Hagimoto et al. in prep.) が目標の25 dBには辿り着いていない
% 今後、FINER受信機との結合試験を行い、FINERシステムで25 dBの達成を目指す

% Note: If compiling with LaTeX+dvipdf, please ensure images generated from 
% other software packages have their bounding boxes set correctly.
%    \begin{figure} [ht]
%    \begin{center}
%    \begin{tabular}{c} %% tabular useful for creating an array of images 
%    \includegraphics[height=5cm]{mcr3b.eps}
%    \end{tabular}
%    \end{center}
%    \caption[example] 
% %>>>> use \label inside caption to get Fig. number with \ref{}
%    { \label{fig:example} 
% Figure captions are used to describe the figure and help the reader understand it's significance.  The caption should be centered underneath the figure and set in 9-point font.  It is preferable for figures and tables to be placed at the top or bottom of the page. LaTeX tends to adhere to this standard.}
%    \end{figure} 

\appendix    %>>>> this command starts appendixes

\acknowledgments % equivalent to \section*{ACKNOWLEDGMENTS}  
We acknowledge K. Iwai for giving us kind advice on our laboratory experiments. 
We also thank S. Nakano, Y. Saito, S. Hotta, J. Yamanaka, S. Inui, M. Kato, M. Sakakibara, Y. Fukuma, S. Fujisawa, and M. Hoshi for their support.
This work was supported by Japan Society for the Promotion of Science (JSPS) KAKENHI Grant Nos. 22H04939, 22J21948, 22KJ1598.
This work was also supported by NAOJ Research Coordination Committee, NINS (NAOJ-RCC-1901-0101, 2001-0104).
 
% This unnumbered section is used to identify those who have aided the authors in understanding or accomplishing the work presented and to acknowledge sources of funding.  

% References
% \bibliography{report} % bibliography data in report.bib
\bibliography{Bibliography} % bibliography data in report.bib

\begin{thebibliography}{10}

\bibitem{Hughes2010}
{Hughes}, D.~H., {J{\'a}uregui Correa}, J.-C., {Schloerb}, F.~P., {Erickson}, N., {Romero}, J.~G., {Heyer}, M., {Reynoso}, D.~H., {Narayanan}, G., {Perez-Grovas}, A.~S., {Souccar}, K., {Wilson}, G., and {Yun}, M., ``{The Large Millimeter Telescope},'' in [{\em Ground-based and Airborne Telescopes III}{\nolinebreak\hspace{0.1em}]},  {Stepp}, L.~M., {Gilmozzi}, R., and {Hall}, H.~J., eds., {\em Society of Photo-Optical Instrumentation Engineers (SPIE) Conference Series} {\bf 7733},  773312 (July 2010).

\bibitem{Hughes2020}
{Hughes}, D.~H., {Schloerb}, F.~P., {Aretxaga}, I., {Castillo-Dom{\'\i}nguez}, E., {Ch{\'a}vez Dagostino}, M., {Col{\'\i}n}, E., {Erickson}, N., {Ferrusca Rodriguez}, D., {Gale}, D.~M., {G{\'o}mez-Ruiz}, A., {Hern{\'a}ndez Rebollar}, J.~L., {Heyer}, M., {Lowenthal}, J., {Monta{\~n}a}, A., {Moreno Nolasco}, M.~E., {Narayanan}, G., {Pope}, A., {Rodr{\'\i}guez-Montoya}, I., {S{\'a}nchez-Arg{\"u}elles}, D., {Smith}, D., {Souccar}, K., {de la Rosa Becerra}, M.~V., {Wilson}, G.~W., and {Yun}, M.~S., ``{The Large Millimeter Telescope (LMT) Alfonso Serrano: current status and telescope performance},'' in [{\em Ground-based and Airborne Telescopes VIII}{\nolinebreak\hspace{0.1em}]},  {Marshall}, H.~K., {Spyromilio}, J., and {Usuda}, T., eds., {\em Society of Photo-Optical Instrumentation Engineers (SPIE) Conference Series} {\bf 11445},  1144522 (Dec. 2020).

\bibitem{Kojima2017}
{Kojima}, T., {Kroug}, M., {Uemizu}, K., {Niizeki}, Y., {Takahashi}, H., and {Uzawa}, Y., ``{Performance and Characterization of a Wide IF SIS-Mixer-Preamplifier Module Employing High-J c SIS Junctions},'' {\em IEEE Transactions on Terahertz Science and Technology}~{\bf 7},  694--703 (Nov. 2017).

\bibitem{Kojima2020}
{Kojima}, T., {Uemizu}, K., {Kiuchi}, H., {Tamura}, T., {Kaneko}, K., {Sakai}, R., {Miyachi}, A., {Shan}, W., {Uzawa}, Y., {Gonzalez}, A., {Kroug}, M., and {Sakai}, T., ``{Wideband technology development to increase the RF and instantaneous bandwidth of ALMA receivers},'' in [{\em Millimeter, Submillimeter, and Far-Infrared Detectors and Instrumentation for Astronomy X}{\nolinebreak\hspace{0.1em}]},  {Zmuidzinas}, J. and {Gao}, J.-R., eds., {\em Society of Photo-Optical Instrumentation Engineers (SPIE) Conference Series} {\bf 11453},  114530P (Dec. 2020).

\bibitem{Carpenter2022}
Carpenter, J., Brogan, C., Iono, D., and Mroczkowski, T., ``The {ALMA2030} wideband sensitivity upgrade,'' {\em arXiv [astro-ph.IM]}  (Oct. 2022).

\bibitem{Klein2012}
{Klein}, B., {Hochg{\"u}rtel}, S., {Kr{\"a}mer}, I., {Bell}, A., {Meyer}, K., and {G{\"u}sten}, R., ``{High-resolution wide-band fast Fourier transform spectrometers},'' {\em \aap}~{\bf 542},  L3 (June 2012).

\bibitem{Iwai2017}
{Iwai}, K., {Kubo}, Y., {Ishibashi}, H., {Naoi}, T., {Harada}, K., {Ema}, K., {Hayashi}, Y., and {Chikahiro}, Y., ``{OCTAD-S: digital fast Fourier transform spectrometers by FPGA},'' {\em Earth, Planets and Space}~{\bf 69},  95 (July 2017).

\bibitem{Finger2015}
{Finger}, R., {Mena}, F.~P., {Baryshev}, A., {Khudchenko}, A., {Rodriguez}, R., {Huaracan}, E., {Alvear}, A., {Barkhof}, J., {Hesper}, R., and {Bronfman}, L., ``{Ultra-pure digital sideband separation at sub-millimeter wavelengths},'' {\em \aap}~{\bf 584},  A3 (Dec. 2015).

\bibitem{Rodriguez2018}
{Rodriguez}, R., {Finger}, R., {Mena}, F.~P., {Alvear}, A., {Fuentes}, R., {Khudchenko}, A., {Hesper}, R., {Baryshev}, A.~M., {Reyes}, N., and {Bronfman}, L., ``{Digital compensation of the sideband-rejection ratio in a fully analog 2SB sub-millimeter receiver},'' {\em \aap}~{\bf 619},  A153 (Nov. 2018).

\bibitem{Whitney2009}
{Whitney}, A., {Kettenis}, M., {Phillips}, C., and {Sekido}, M., ``{VLBI Data Interchange Format (VDIF) (invited)},'' in [{\em 8th International e-VLBI Workshop}{\nolinebreak\hspace{0.1em}]},   42 (Jan. 2009).

\bibitem{Tamura2020}
{Tamura}, Y., {Kawabe}, R., {Fukasaku}, Y., {Kimura}, K., {Ueda}, T., {Taniguchi}, A., {Okada}, N., {Ogawa}, H., {Hashimoto}, I., {Minamidani}, T., {Kawaguchi}, N., {Kuno}, N., {Togami}, Y., {Hagimoto}, M., {Nakano}, S., {Matsuda}, K., {Okumura}, S., {Nakamura}, T., {Kurita}, M., {Takekoshi}, T., {Oshima}, T., {Onishi}, T., and {Kohno}, K., ``{Wavefront sensor for millimeter/submillimeter-wave adaptive optics based on aperture-plane interferometry},'' in [{\em Ground-based and Airborne Telescopes VIII}{\nolinebreak\hspace{0.1em}]},  {Marshall}, H.~K., {Spyromilio}, J., and {Usuda}, T., eds., {\em Society of Photo-Optical Instrumentation Engineers (SPIE) Conference Series} {\bf 11445},  114451N (Dec. 2020).

\bibitem{Riley2008}
Riley, W. and Howe, D., ``Handbook of frequency stability analysis,'' (2008-07-01 00:07:00 2008).

\end{thebibliography}
\bibliographystyle{spiebib} % makes bibtex use spiebib.bst

\end{document}